\documentclass{IEEEconf}

 \makeatletter
 \def\ps@headings{%
 \def\@oddhead{\mbox{}\scriptsize\rightmark \hfil \thepage}%
 \def\@evenhead{\scriptsize\thepage \hfil \leftmark\mbox{}}%
 \def\@oddfoot{}%
 \def\@evenfoot{}}
 \makeatother
\usepackage{latex8}
\usepackage{times}

\makeatletter
\newif\if@restonecol
\makeatother

\usepackage[]{algorithm2e} 

\usepackage{cite}
\usepackage{url}
\usepackage{graphicx}
\usepackage{subfigure}
\usepackage{balance}
\usepackage{amssymb}
\usepackage{amsmath}
\usepackage{graphicx}
\usepackage{comment} 
\usepackage{amsfonts} 
\usepackage{amsmath} 
\usepackage{rotating}

\usepackage{balance}
\usepackage{subfigure}

\usepackage{amssymb}
\usepackage{amsmath}
\usepackage{graphicx}
\usepackage{comment} 
\usepackage{amsfonts} 
\usepackage{mathptm}

\newtheorem{theorem}{Theorem}[section]

\newtheorem{definition}[theorem]{Definition}

\begin{document}

\title{The Smallville Effect: Social Ties Make Mobile Networks\\ More Secure Against the Node Capture Attack}

\author{
        \normalsize
         \parbox{2 in}{\centering Mauro Conti   \thanks{Also with Center for Secure Information Systems, George Mason University, Fairfax, VA, USA. E-mail: mconti1@gmu.edu.}
         \\
         \emph{Department of Computer Science}\\
         \emph{Vrije Universiteit Amsterdam}\\
        \emph{1081 HV - Amsterdam, \\The Netherlands}\\
        \emph{mconti@few.vu.nl}\\
        }
         \hspace*{ 0.1 in}
         \parbox{2 in}{ \centering Roberto Di Pietro\thanks{Also with Dipartimento di Matematica, Universit\`a di Roma Tre, Roma, Italy. E-mail: dipietro@mat.uniroma3.it.}         \\
         \emph{UNESCO Chair in Data Privacy}\\
        \emph{Universitat Rovira i Virgili}\\
        \emph{43700 - Tarragona, Spain}\\
        \emph{roberto.dipietro@urv.cat}\\
}
         \hspace*{ 0.1 in}
         \parbox{2.2 in}{ \centering Andrea Gabrielli,\\ Luigi V. Mancini, \\and Alessandro Mei\\
         \emph{Dipartimento di Informatica}\\
         \emph{Universit\`a di Roma ``Sapienza'' }\\
        \emph{00198 - Roma, Italy}\\
        \emph{\{a.gabrielli, mancini, mei\}@di.uniroma1.it}\\
        }
}

\maketitle

\begin{abstract}
Mobile Ad Hoc networks, due to the unattended nature of the network itself and the dispersed location of nodes, are subject to several unique security issues. One of the most vexed security threat is node capture. A few solutions have already been proposed to address this problem; however, those solutions are either centralized or focused on theoretical mobility models alone. In the former case the solution does not fit well the distributed nature of the network while, in the latter case, the quality of the solutions obtained for realistic mobility models severely differs from the results obtained for theoretical models. The rationale of this paper is inspired by the observation that re-encounters of mobile nodes do elicit a form of social ties. Leveraging these ties, it is possible to design efficient and distributed algorithms that, with a moderated degree of node cooperation, enforce the emergent property of node capture detection.
In particular, in this paper we provide a proof of concept proposing a set of algorithms that leverage, to different extent, node mobility and node cooperation---that is, identifying social ties---to thwart node capture attack. In particular, we test these algorithms on a realistic mobility scenario. Extensive simulations show the quality of the proposed solutions and, more important, the viability of the proposed approach.

\end{abstract}


\section{Introduction}
\label{introduction}

Ad Hoc networks are an ideal candidate for the deployment in harsh environments,
due to their capacity of operating without an existing infrastructure. The application
scenarios include law enforcement, search-and-rescue, disaster recovery, and others.
In these cases, Ad Hoc Networks have the additional appealing feature to be able to operate in
an unattended manner. However, this comes at a cost: Ad Hoc networks are vulnerable to different kinds of novel attacks. For instance, an adversary could eavesdrop all the network communications, or it might capture (i.e. remove) nodes from the network.
Captured nodes can be re-programmed and re-deployed in the network area, with the goal of subverting the data aggregation, the decision making process, or other operations.
Moreover, they can be re-programmed, replicated in many copies, and re-deployed
in the network to perform any sort of vicious attack, amplified by the presence of many
malicious devices. In this paper, we start from the observation that all of these
attacks start from the capture of one of the nodes. Therefore, being able to detect
this malicious activity becomes a formidable way to stop many of the threats to
Ad Hoc networks.

The node capture attack may also be of independent interest, an example comes
from the LANdroids~\cite{landroid} research program by the  U.S. Defense Advanced Research
Projects Agency (DARPA). This research program has the goal of developing
smart robotic radio relay nodes for battlefield deployment. LANdroid mobile nodes are supposed
to be deployed in a hostile environment, establish an ad-hoc network, and provide connectivity
as well as valuable information for soldiers that would later approach the deployment area.
An adversary might attempt to capture one of these nodes to reduce the efficiency of
the network.

The unique requirements of the Ad Hoc network context call for efficient and distributed solutions to the node capture attack.
We believe that any solution to this problem has to satisfy the following requirements:
(i) to detect the node capture as early as possible; (ii) to have a low rate of false positives---nodes that are believed to be captured and thus subject to a revocation process, but that were not actually taken by the adversary; (iii) to have a low rate of false negatives--nodes that are believed to be not captured, but that were actually captured; (iv) to introduce a small overhead.
The solutions proposed so far are not efficient \cite{PSW:SecurityWSN:CommunicationACM}. Moreover, na\"{i}ve centralized solutions, although they can in principle be applied, present drawbacks like single point of failure and non uniform energy consumption.

In this paper we tackle the problem of detecting the node capture attack in the context of mobile
Ad Hoc networks of devices like smart-phone or PDAs carried by individuals.
These networks have attracted the
attention of a large number of researchers in the field of networking. They
are often called Pocket Switched Networks, and they are part of the class of the Delay Tolerant
Networks. In Pocket Switched Networks the contacts between devices are used as
opportunities of message forwarding. Actually, a large part of the work in this field has
the goal of designing multi-hop routing mechanisms that are able to deliver messages
between any arbitrary pair of devices in the network efficiently.
In these networks the mobility pattern has some unique features. Since devices
are carried by humans, the pattern of contacts between devices mimic the social nature
of human mobility. This observation has been used to build forwarding mechanisms
that use the notion of community to make messages find their way from source to
destination.\\
In particular, this paper describes how to make use of the social nature of contacts to get
a network stronger against the node capture attack. To the best of our knowledge,
this is the first work that demonstrates that social-based mobility has a strong impact
on security. In our solution, nodes are responsible, in a distributed fashion, to monitor
the presence of one or more other peers. If the mobility pattern were irregular, or
arbitrary, or if every node moved independently, like in the random way-point
mobility model, then any choice of pairs of monitoring and monitored node would
essentially be the same. Our key observation is that social mobility has a unique pattern, and so
there is a way to assign responsibilities in the network to improve
considerably the efficiency of our protocols for the node capture attack.
Quite naturally, we will see that the the best performance is achieved when the
monitoring node and the monitored node have a strong social tie. Just like
in the real life, if we disappear for some time the first persons that get worried are
our family members and our friends at work. Or just like what happens in small villages,
where social ties are traditionally very strong and where,
if something wrong happens, very quickly someone realizes and alert everybody.
This is why we call this phenomenon the Smallville effect.

We will describe the problem in terms of capture and revocation of captured nodes. However, our
solution can be applied also in more general scenarios. As an example, there could be a trust relationship established between a group of users such that they stop trusting nodes that disappear
from the network for a given time-interval. The node, in case its absence was not due to
a capture, could be asked to go through some expensive and secure procedure (e.g. obtaining a new communication key form a central server) in order to join again the group in the
trust-relationship.

We will validate our solution with a large set of experiments performed using real traces
publicly available, those collected at INFOCOM 2005, and we
will see that the Smallville effect can be considerably strong.

The rest of the paper is organized as follows. In Section \ref{Related Work} we review the related work in the area. In Section \ref{assumptions} we present the system model and the assumption used in this work.
Our proposal is presented in Section \ref{theprotocol}, while the performances results are discussed in Section \ref{Simulations}. We conclude the work in Section \ref{conclusions}.

\section{Related Work}
\label{Related Work}
Wireless social community networks are emerging as an alternative to traditional networks to provide wireless data services. This type of networks relies on users---a wireless community can rapidly deploy a high-quality data access infrastructure in an inexpensive way.
To the best of our knowledge, no security issues have been investigated in the particular context of social wireless communities networks. Interest in wireless social communities network has been recently shown by the research community from different point of views, like network coverage \cite{SocialCommunitiesINFOCOM2008} or the detection of the source-starvation \cite{INFOCOM2008:starvation}. Also the identification of communities in more traditional network, such as the file sharing peer-to-peer network, has been of interest for the research community \cite{Small-worldFileSharing}.

Mobility as a means to enforce security in mobile networks has been considered in \cite{Capkun_Hubaux:MobilityHelpsSecurity:MobiHoc:2003}. 
In \cite{WiSec08:Perrig:MindYourManners}, the authors identified social and situational factors which
impact group formation for wireless group key establishment.
Further, mobility has been considered in the context
of routing~\cite{Grossglauser:humanspeed:INFOCOM2003} and of network property optimization~\cite{Hubaux:JointMobilityAndRouting:INFOCOM2005}. 
In particular, \cite{Grossglauser:humanspeed:INFOCOM2003} leverages node mobility in order to disseminate information about destination location without incurring any communication overhead. In \cite{Hubaux:JointMobilityAndRouting:INFOCOM2005} the sink mobility is used  to optimize the energy consumption of the whole network. A mobility-based solution for detecting the sybil attack has been recently presented in \cite{Piro:DetectingSybilInMobile_SecureComm2006}.
Finally, note that a few solutions exist for node failure detection in ad hoc networks \cite{HsinLiu:SelfMonitoring:2006, HsinLiu:ADistributedMonitoring:WISE02, Ranganathan:Gossip-Style:ClusterComputing:2001, Hay:Failure:SRDS02}. However, such solutions assume a static network, missing a fundamental component of our scenario, as shown in the following.

Node capture attack is considered as major threat in many security solutions for WSN.
In particular, in \cite{Huang:SASN04} both oblivious and smart node capture is considered for the design of a key management scheme for WSN. A deeper analysis on the modeling of the capture attack has been presented in \cite{Tague:AdHocElsevier:2007, Tague:Vulnerability:INFOCOM2008}. In \cite{Tague:AdHocElsevier:2007}, it is shown how different greedy heuristics can be developed for node capture attacks and how minimum cost node capture attacks can be prevented in particular setting. In \cite{Tague:Vulnerability:INFOCOM2008}, the authors formalize node capture attacks using the vulnerability metric as a nonlinear integer programming minimization problem.

Node mobility and node cooperation in a mobile ad hoc setting has been considered already in Disruption Tolerant Networks (DTNs)~\cite{Sterbenz:Survivable(DTN):WISE2002, Elizabeth:SocialNetwork:MobiHoc07}. However, such a message passing paradigm has not been used, so far, to support security. We leverage the concept introduced with DTN to cooperatively control the presence of a network node.

In \cite{CDMM:WiSec08:Emergent, CDMM:EURASIP:2009} a proof of concept that it is possible to design a node capture detection protocol leveraging the network mobility is given---more specifically leveraging the expected ``re-meeting'' time between nodes.
However, \cite{CDMM:WiSec08:Emergent, CDMM:EURASIP:2009} present capture detection solutions focusing on a specific mobility model, the Random Waypoint Mobility (RWM) model \cite{Broch:RWM:MobiCom98}.
RWM shown different problems. One of these is that the average speed of the network tends to decrease during the life of the network itself and, if the minimum speed that can be selected by the nodes is zero, then average speed of the system converges to zero~\cite{yoon03random}.
In \cite{yoon03random} it is also suggested to set the minimum speed to a value strictly greater than zero. In this case, the average speed of the system continue decreasing, but it converges to a non-zero asymptotic value. Other problems related to spatial node distribution have been considered by different authors~\cite{Esa:SpatialDistributionRWM,yoon03random, Resta:AnAnalysisOfTheNode_POMC02}.
Finally, the RWM model can be far from describing realistic mobility patterns \cite{Esa:SpatialDistributionRWM,yoon03random, CDGMM:WWIC:TheQuest}. The work in \cite{CDGMM:WWIC:TheQuest} highlighting that in realistic mobility:
\begin{enumerate}
\item Some single nodes meet all the other nodes with a very low frequency. In the following we will refer to such a type of node as to \textit{isolated} nodes.
\item There are subset of nodes that meet between them with a significantly higher frequency than the average. We will refer to such a type of subset as to \textit{communities}. As examples of everyday life, we can think to students that attend the same class or people that work or live in the same building.
\end{enumerate}

In this work we do not consider mobility traces synthesized using the RWM model.
Instead, we consider only real traces.
Among the publicly available traces for mobile nodes (e.g. from \cite{CRAWDAD}), we consider the traces collected at INFOCOM 2005 conference \cite{CRAWDAD-INFOCOM}, already used in previous research works \cite{Hui:WDTN05,Chaintreau:CoNEXT07,Chuah:Simplex09,Tang:WOSN09}. In particular, the traces of mobile nodes were gathered using Bluetooth devices distributed to 41 people attending the INFOCOM 2005 conference.

\section{System Model and Assumptions}
\label{assumptions}

In the following we state all the assumptions used throughout this paper as well as the overhead model used to assess the performance of the proposal.

First, we clearly state the threat we are going to address.
\begin{definition}
\textbf{Node Capture.}
An adversary physically removes a node from the network---or just tamper with the node---in a way such that the node can not communicate with the other nodes in the network. The attack can last forever or just for a given period of time.
\end{definition}

Table \ref{tabella_notazione} resumes the notation used in this paper.

{\bf Network assumptions}
We assume that the system provides a broadcasting primitive. This primitive is easily
implementable by using a flooding mechanism.
Further, we assume that nodes in the systems have a protocol abiding behavior, and that no compromised node is present when the protocol starts---indeed, node capture is a pre-requisite for node compromise.

Moreover, in the proposed solution every node maintains its own clock. To embrace the more general case, we also assume that nodes are not equipped with localization  devices, like GPS.
However, we require that clocks among nodes are loosely synchronized. Note that there are a few solutions proposed in the literature to provide loose time synchronization, like \cite{wsn_secure_synchro}.
Therefore, in the following we will assume that skew and drift errors are negligible.
Finally, security of network communications is out of the scope of this work; however, note that in order to enforce required security properties, we could rely on one of the many protocols proposed in the literature.
For instance, the solution in \cite{winet} could be used to provide node authentication.

{\bf Message overhead model} The main overhead introduced by the protocol due to message broadcast.
In \cite{WC:ComparisonBroadcasting:MobiHoc02} a classification of the different solutions for broadcasting scheme is provided:
(i) Simple Flooding; 
(ii) probabilistic-based schemes;
(iii) area based schemes that assume location awareness;
(iv) neighbor knowledge schemes that assume knowledge of two hop neighborhood.
Analyzing or comparing broadcasting cost is out of the scope of this paper. However, for a better comparison of the solutions proposed in this paper, we need to set a broadcast cost that will be expressed in terms of unicast messages. In fact, the overhead associated to the broadcasting varies with different network parameters (for instance, node density and communication radius).
A deeper analysis on the overhead generated  for different broadcasting protocols is presented in~\cite{OPPV:LocaTechn:2004}.

Finally, note that a message could be received more than once, for instance because the receiver is in the transmission range of different rely nodes.
However, in the following we assume that a broadcasted message reaching all the nodes is received (then counted) only once for each node---it costs as 1 sent and 1 received message for each node.
A similar assumption is used for example in~\cite{OPPV:LocaTechn:2004}.

\begin{table}
\begin{center}
\begin{footnotesize}
\caption{Time-related notation}
\vspace{0.3cm}
\centering
\begin{tabular}{cl}
\hline
\textbf{Symbol} & \textbf{Meaning} \\
    \hline\hline
$\sigma$ & Message propagation delay. \\
    \hline
$\tau$ & Interval time between presence claim\\
       & for the Benchmark protocol. \\
    \hline
$\lambda$ & Alarm time-out.\\ 
    \hline
$\delta$ & Time available to the allegedly captured node \\
         & to prove its presence.  \\ 
\hline
$\gamma$ & Interval time for node cooperation requests \\
	& in the AdaBo protocol. \\
\hline
$n$ & Number of nodes in the network. \\
    \hline
$K$ & Total number of nodes tracked by each node\\
    \hline
$K_A$ & Number of nodes tracked by each node\\
      & using adaptiveness.\\
    \hline
$K_B$ & Number of nodes tracked by each node\\
      & using booking.\\
    \hline
\end{tabular}
\label{tabella_notazione}
\end{footnotesize}
\end{center}
\end{table}

\section{The Protocol}
\label{theprotocol}

In this section, we present our solution.
In particular, to help the reader capture the insights of our final proposal,
we refer to the case where just one node is captured and describe in the following two reference solutions:
\begin{itemize}
\item \textit{Benchmark protocol.} The benchmark protocol is a simple solution that does not use node mobility and contact patterns~\cite{CDMM:EURASIP:2009}. 
We briefly report this solution in Section \ref{benchmark}.
\item \textit{Base protocol.} Similarly, the
 {\em base} protocol introduced in~\cite{CDMM:WiSec08:Emergent, CDMM:EURASIP:2009}
is briefly recalled in Section \ref{base}.
\end{itemize}

We use both the Benchmark and the Base protocol as simple reference protocols to compare with.
After these simple solutions, we present other two protocols, each of them capturing different aspects of the realistic mobility model introduced:
\begin{itemize}
\item \textit{Booking protocol.} This protocol addresses the \textit{isolated} nodes that can result in
realistic mobile environments. It is described in Section \ref{booking}.
\item \textit{Adaptive protocol.} This protocol leverages the \textit{communities} that naturally emerge in realistic mobile environments. It is described in Section \ref{adaptiveProtocol}.
\end{itemize}

Finally, we combine the two previous protocols as building blocks of our final proposal, the {\em AdaBo} protocol.

\subsection{Benchmark Protocol}
\label{benchmark}

In this section, we report a na\"{i}ve solution for the node capture detection that does not make
use of node mobility. First, assume that a Base Station is present in the network---we will show later how to remove this assumption. Each node periodically (for instance, every $\tau$ seconds) sends a message to the BS carrying  some evidence of its own presence. In this way, the base station can check whether a node is present.
If a node does not send the claim of its presence to the BS when it is assumed to do that (after $t$ seconds from the previous claim), the base station will revoke the corresponding node ID from the network  (for instance, flooding the network with a revocation message).

To remove the centralization point given by the presence of the BS, we require each node to notify its presence to any other node in the network (instead of just to the BS). A node can prove its presence throughout a broadcasted and flooded message.
A node receiving this claim would restart a time-out set to $\tau+\sigma$, where $\sigma$ accounts for network propagation delay. Should the presence claim not be received before the time-out elapses, the revocation procedure would be triggered. However, note that if a node is required to store the ID of any other node as well as the receiving time of the received claim message, $O(n)$ memory locations would be needed in every node.
To reduce the memory requirement on node, it is possible to assume that the presence in the network of each node is tracked by a small subset of the nodes of the network. Hence, if a node is absent from the network for more than $\tau$ seconds, its absence can still be detected by a set of nodes.

Note that for the Benchmark protocol the average number of messages $m(t_u)$ each node sends (the equation actually holds also for the number of received messages) over a time unit $t_u$ obeys to the following equation:
\begin{equation}
m(t_u)=\frac{t_u}{\tau}\cdot n
\end{equation}
where $n$ is the number of nodes, and $\tau$ is the time-interval between presence floodings sent by a single node. Assuming a smart attacker that captures the node $a$ just after the presence claim flooded by node $a$, $\tau$ also corresponds to the detection time.

\subsection{Base Protocol}
\label{base}

In this section, we report a recent  protocol  \cite{CDMM:WiSec08:Emergent, CDMM:EURASIP:2009} designed for the RWM model \cite{Broch:RWM:MobiCom98}. In the present work, we refer to this previous proposal as the {\em base} protocol.
The approach in \cite{CDMM:WiSec08:Emergent, CDMM:EURASIP:2009} is based on the following observation. First, if node $a$ has eavesdropped a transmission originated by node $b$, at time $t$, we will say that a {\em meeting} occurred. Now, nodes  $a$ and $b$ are mobile, so they will leave the communication range of each other after some time. However, these two nodes are expected to re-meet again within a certain time-interval, or at least with a certain probability within a certain time-interval.

In the Base protocol, each node $a$ is given the task of witnessing for the presence of a specific set $T_a$ of other nodes (we will say that $a$ is  \emph{tracking} nodes in $T_a$). In particular, the node $a$ selects the nodes to be in $T_a$ as the first $K$ nodes $a$ meets (where $K$ is the desired cardinality of $T_a$).
For each node $b \in T_a$ that $a$ gets into the communication range of, $a$ sets the corresponding meeting time to the value of its internal clock and starts the corresponding time-out, that would expire
after $\lambda$ seconds. As a protocol option, the meeting nodes can also cooperate, exchanging information on the meeting time of nodes of interests---that is, nodes that are tracked by both $a$ and $b$.
If the time-out expires (that is, $a$ and $b$ did not re-meet within $\lambda$ seconds), the network is flooded with a node-missing alarm triggered by node $a$. If node $b$ does not prove its presence within  $\delta$ seconds after the broadcasted node-missing alarm is flooded, every node in the network will revoke node $b$.

\subsection{Booking Protocol}
\label{booking}

\begin{figure}
\begin{center}
\includegraphics[scale=0.30]{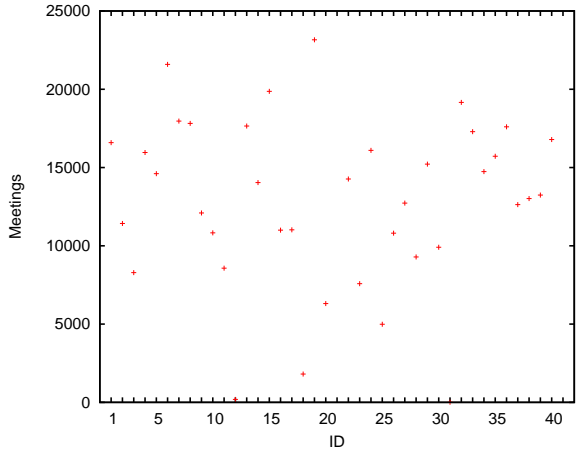}
\end{center}
\caption{\label{INFOCOM:meetings}INFOCOM traces: Nodes meetings}
\end{figure}

In this section, we address the first characterization of realistic mobility model.
That is, there are some \textit{isolated} nodes that meet all the other nodes with a very low frequency (compared to the meeting frequency the other nodes have).
In particular,
an \text{isolated} node can appear for the first time in the network (i.e. having the first meeting with another node) a while after the network operations are started.
Observe that, if such a node is captured before the first meeting, there would be no node tracking it (hence its capture would go undetected), for the Base protocol.
An evidence of the presence of \textit{isolated} nodes can be seen in Figure \ref{INFOCOM:meetings}. In fact, Figure \ref{INFOCOM:meetings} shows for each node (INFOCOM traces) on the x-axis, the number of meetings for that node, on the y-axis. From Figure \ref{INFOCOM:meetings} it is possible to note how node 13, as well as node 18, can be considered as \textit{isolated} nodes.

Leveraging  this observation, we introduce the Booking protocol.
Actually, this protocol is just a slightly modification of the Base protocol. In particular, the only difference is in the way the $K$ nodes that a node is going to track are selected.
In particular, we assume that the network administrator decides for every node what are the IDs of the other nodes it is going to track. In this way, the network administrator can guarantee that every node is tracked by a fixed number of other nodes.
The Booking protocol not only guarantees that every node is going to be tracked by someone (property not satisfied by the Base protocol); it is also possible for a node $a$ to revoke a node $b$ that it never meets (e.g. $b$ could had been captured at the network deployment time or before the node $a$ met it for the first time).

We observe that in the Booking protocol communities are not  leveraged  to optimize the node tracking.

\subsection{Adaptive Protocol}
\label{adaptiveProtocol}

In
realistic mobility pattern, it has been observed \cite{CDGMM:WWIC:TheQuest} that there are subsets of nodes which elements meet between them with higher frequency than the average (here,
\textit{communities} emerge).

Differently from the \textit{base} protocol, that has been designed for the RWM model, we can actually leverage this behavior in the design of the capture detection protocol for a realistic mobile environment.
In fact, we expect that the capture detection protocol would have better performances if we were able to let the nodes track the other nodes that they meet with higher frequencies, instead of just any $K$ out of the $N$ nodes.
In general, this improvement can lead to a more efficient protocol (lower number of node-missing alarm) or to a more effective protocol (lower detection time).

As the communities cannot be always predicted (e.g. by the network administrator), we would also like that the nodes autonomously discover who are the nodes that they meet with higher frequency.
Furthermore, a nice property of the protocol would be for the node to adapt its set of tracked nodes in the case that the mobility pattern changes.

With all these observations as a rationale, we design the Adaptive protocol. The aim of this protocol is that while the time goes by, the $K$ nodes that a node is tracking are the $K$ better nodes for it to be tracked, i.e. out of the $N$ nodes, the $K$ nodes for which it does not raise node-missing alarm.
The behaviour of the Adaptive protocol can be summarized as follow:
\begin{itemize}
\item Node $a$ starts tracking the first $K$ nodes it meets (the first selection of the tracked nodes is as for the Base protocol).
\item When $a$ raises a node-missing alarm for a node, say $b$, while $b$ actually proves it presence, $a$ stops tracking node $b$ and start tracking the next node it meets.
\end{itemize}
As an enhancement to this protocol, we let the nodes using some memory slots to silently track nodes. We will refer to such type of slots as Silent Memory Slots (SMSs). In particular, a node $a$ uses these slots as follows:
\begin{itemize}
\item The SMSs are populated as for the regular tracking slot, considering the IDs of the first nodes that $a$ meets.
\item For each node in the SMSs, node $a$ keeps note of the number of meetings (normalized for a time unit). In the following, we refer to such statistical value as the $score$.
\item When node $a$ raises a node-missing alarm for a node, say $b$, that actually proves its presence, the next node to be tracked is not chosen as the first newly encountered node. Instead, the next node to be tracked is the first in the SMSs list. Note that we consider the SMSs list ordered based on the $score$s. In this way, node $a$ will track the nodes that it met with higher frequency; intuitively, the best one to be tracked among the nodes  $a$  has in the SMSs.
\item At regular time-intervals (a protocol parameter), $a$ removes from its SMSs the node with the lower $score$, and put-in the next node it meets.
\end{itemize}

\subsection{The AdaBo Protocol}
\label{caching}

On one hand, the Booking protocol aims to guarantee that all the nodes (including the \textit{isolated} ones) are tracked. However, its efficiency could be questionable.
On the other hand, the Adaptive protocol aims to let each node be efficiently tracked while not giving the guarantee provided by the Booking protocol. In this section, we describe the AdaBo protocol that gives the guarantee of the Booking protocol---i.e. that all the nodes are tracked---while being quite efficient.
Further, optimization are also introduced.

The simple idea we start from is to dedicate: (i) part of the node's memory slot to be managed according to the Booking protocol; (ii) the remaining portion of the node's memory to managed according to the Adaptive protocol.

\begin{figure}
\begin{center}
    \includegraphics[scale=0.25]{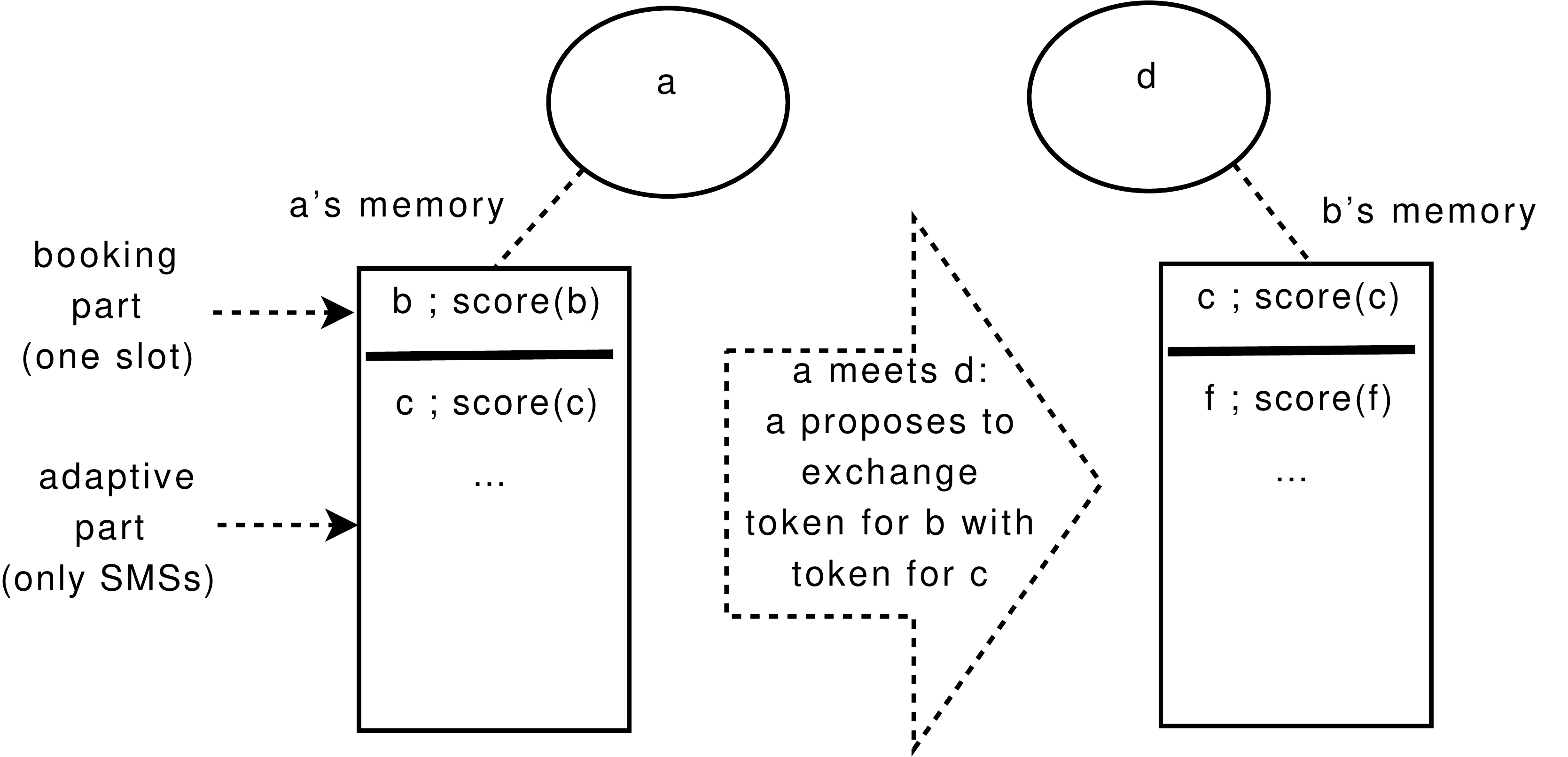}
\end{center}
\caption{AdaBo protocol: booking token exchange proposal.}
\label{exchangeExample}
\end{figure}

As for the number of memory slots to be dedicated to the Booking protocol, we just observe that for each node $a$, having one other node to track it is enough to guarantee that every node is tracked by at least one node. Indeed, for any captured node there will be a node detecting the capture and raising the corresponding node-missing alarm.
As a result, we need that each node tracks in booking mode just one other node to have the above property holding. For ease of exposition, in the results presented in Section \ref{Simulations} we will consider the {\em AdaBo} protocol where just one memory slot is dedicated to the booking approach. 
In this section, we refer to the example of Figure \ref{exchangeExample}.

As pointed out in Section \ref{booking}, the booking approach does not leverage the communities. That is, it does not aim to efficiently assign the tracking of a node $a$, to the best node that can track it---e.g. the node that meets $a$ with higher frequency. We further observed (Section \ref{base}) that for a node $a$, tracking the first node that it meets, is a na\"ive choice to leverage communities. In the AdaBo protocol, we optimize the way the booking part of the memory is used. In particular, all the memory dedicated by the AdaBo to the adaptive approach will be considered as SMSs. The information about nodes in SMSs will be used to optimize the booking part---allowing nodes to exchange the nodes they are tracking in booking. AdaBo protocol can be described as follows:

\begin{itemize}

\item \textit{Initialization.} First, we let each node start having in booking the token---i.e. the ID---of himself. Of course, this is done just as a set-up choice. The node will not actually track himself---there would be no utility in doing it as the node should detect the capture of himself. Furthermore, we assign to each node a maximum number of available token exchanges it can be part of. When the protocol starts each node has not participated in any exchange yet.

Each node uses all the memory dedicated to the adaptive approach in the silent mode (Section \ref{adaptiveProtocol})---i.e. just keeping note of the number of meeting over a time unit for the nodes in this memory.

\item \textit{Start-up.} We give to the nodes a set-up time-interval (that is a protocol parameter) during which the nodes just collect statistics of the meetings with other nodes (for the $score$ in the SMSs). No node-missing alarms are raised in this time-interval; no token exchanges are made as well.
As soon as the set-up time expires. Each node will try to pass-it-on the token it holds (referred to himself) to the node in the SMSs with higher $score$---the node he met more frequently. The token exchanges occurs based on node meeting and one-hop communication only. 
That is, in this case the exchange will be done as soon as the node meets its first node in the SMSs, $c$.
Note that in this case the node's aim is to be tracked by some other node.

\item \textit{Iterated step.} After the first token exchange (previous point), the node, say $a$, continues improving the quality of its tracking; that is, to track the best node it can track, from the $score$ point of view---let $c$ be this node.
Note that, as a consequence of the previous exchanges, this condition can not yet hold for all the nodes just after the first exchange. Furthermore, this condition can also be unachievable for some node $a$: this happens either if (i) $a$ never meets the node that is having $c$ in booking; or, (ii) the node that is tracking $c$ in booking is not available for the token-exchange, accordingly with the evaluation described in the following.
Despite this, the node $a$ tries to reach its final (potential) target (tracking the node with the highest $score$ in its SMSs) in a greedy way:
as soon as it has the chance to improve the quality of its tracking, $a$ attempt to improve it. For example, assume node $a$ meets node $d$,
that it is currently tracking node $c$.
Assume that $a$ checks that it can track the node $c$ with higher performance when compared to the quality of the tracking of node $b$ that is currently booking ($a$ checks that it is having with $c$ an higher number of meetings compared to the meetings with $b$). In this case, node $a$ proposes to the node $d$ to exchange the tokens---i.e. the IDs---of the nodes they are respectively tracking in booking. Whether the exchange actually succeeds depends
on node $d$. In particular, on whether node $d$ will decrease the performance over the booked node. To evaluate this, node $d$ checks, for the currently booked node $c$, its $score$ value (if any; a node in booking is not necessary also in the SMSs). If a node proposed for the exchange ($c$ or $b$) is not in $d$'s SMSs, it is assigned $score = -\infty$. Finally, node $d$ will agree for the exchange iff: $$score(c)\ge score(b).$$ Note that, if node $d$ does have neither  $c$  nor $b$ in its SMSs, the previous equation accounts for $d$ helping $a$ improving its tracking performances.
Finally, observe that in any case an exchange happens, both exchanges counter of $a$ and $d$ are incremented. Also, note that if a node reached the maximum number of exchanges it can participate in, it will not be able to propose exchanges.
\end{itemize}

In the AdaBo protocol we also consider an improvement on the way a node sends the node-missing alarm. That is, assume a time-out $\lambda$ relative to node $b$ is expiring on node $a$. Node $a$, before flooding the network with a node-missing alarm for $b$, will ask the nodes it meets in the last small time-interval $\gamma$ of $\lambda$, if they can prove him the  presence of $b$ within the last $\lambda$ time-interval. If the latter is the case, $a$ will update $b$'s presence with the time just proved to him. Otherwise, as it happens for all the other  presented protocols, $a$  will flood the network with a node-missing alarm relative to the capture of node $b$.

We observe that for the described solution the following problems could arise. If a node is captured while it has the token of himself, no one will detect its capture.
Such a problem could be solved by considering one more booking slot (used only for the set-up time-interval) where IDs in booking are assigned by the network administrator in such a way that each node does not have its own token. After the set-up time---i.e. after the nodes have given  the token referred to themselves to some other nodes---this second booking slot can be used as SMSs.

A similar problem can also be found if we want to deploy more nodes in subsequent times. A solution similar to the one just described for the initial phase can be used. In this case, an undesirable property will hold: the new set of deployed nodes would be considered as an independent network itself. That is, at the time of the new deployment, a node belonging to the set of newly deployed nodes will be tracked just by nodes from the same set. However, note that as soon as the newly deployed nodes meet the previously deployed ones, the former ones will exchange tokens with the latter ones, so removing the initial undesired property.
However, if a single alone node is deployed, we should use some other mechanism; e.g. the BS could communicate with a node already in the network to exchange the booking token with the newcomer.

Finally,
note that we described our solution privileging ease of exposition.
 However, the proposed solution could also address the problem of the changes in mobility patterns.
Such pattern mobility changes can be captured by the proposed protocol re-running the start-up phase at regular time-interval.
We leave as future work a detailed investigation of this issue, together with an assessment of the efficiency of such mechanisms.

\section{Simulations and Discussion}
\label{Simulations}

In this section, we present the results of the simulations that we made in order to asses the performance of the proposed solution. In particular, the main aim of the simulation has been to investigate the protocol effectiveness (i.e. the detection time) versus the protocol efficiency (i.e. the cost in terms of messages). Furthermore, we also investigated the false negative rate of the different protocols presented in Section~\ref{theprotocol}.
We point out that our protocols do not have false positives. In fact, assume that node $a$ floods the network with a node-missing alarm related to node $b$ while $b$ is actually within the network---we assume in this case $b$ can communicate with the other network nodes. In our protocols, $b$ has $\delta$ seconds to prove its presence: $\delta$ accounts for the propagation time of both (i) the node-missing message sent by $a$; and, (ii) the presence proving message sent by $b$. Hence, on the one hand, if node $b$ can not be reached by the flooded node-missing alarm it means $b$ is isolated from the other nodes: correctly considered as captured, indeed.
On the other hand, if it is reached by the node-missing alarm, it can prove its presence---preventing false positives to occur.


We implemented a simulator of our protocols that takes as input a trace of nodes mobility---every nodes meeting is described by the couple of participating nodes ID and the time of the meeting.
We ran multiple simulations of the protocols we proposed in Section~\ref{theprotocol}. We used as an input traces derived from the mobility traces collected at INFOCOM 2005 \cite{CRAWDAD-INFOCOM}. We describe the traces in Section~\ref{the-infocom-trace}. The setting of the simulation are described in Section \ref{simSetting}. Finally, the simulation results are presented in Section \ref{OVperformances}.

\subsection{Real Traces}
\label{the-infocom-trace}

The traces considered in our simulation have been obtained from
the mobility traces collected during the INFOCOM 2005 Conference \cite{CRAWDAD-INFOCOM}. Information for the traces in \cite{CRAWDAD-INFOCOM} has been gathered using Bluetooth devices distributed to 41 people attending the conference. In particular, we are interested in the mobility and the social interaction between people during the daylight. Thus, we selected from these traces the events within the 73,000th second and the 115,000th second. Then, we removed:  (i) the events related to Bluetooth devices not explicitly involved in the experiment (that are, devices not assigned by the experiment organizer \cite{CRAWDAD-INFOCOM}); and, (ii) events related to devices involved in the experiment but not reporting any meeting in the selected time-interval (node with IDs 21 and 41). The resulting traces have 39 nodes.
To run extensively simulation, we considered 10 times the subsequent repetition of the obtained events. This choice is motivated by the fact that nodes of social networks tend to repeat their mobility pattern.
The resulting traces consists of 420,000 seconds of events.
Finally, we assume that the events in the traces are symmetric: if node $a$ meets node $b$ (that is node $a$ knows to be in the communication range of node $b$), then node $b$ meets $a$ too. Observe that the resulting traces we used in our simulation still maintain the characteristics of a social network: it shows a power-law inter-meeting time. 

\subsection{Simulation Setting}
\label{simSetting}
We simulated different node captures, varying the captured node and the capture time. In particular, starting from 100,000 seconds after the network deployment, we have considered the events split in 13 intervals of 6 hours each. For each of the 39 nodes in the traces, we simulated the capture at the beginning of each of these intervals. In other words, the first capture is simulated at 100,000 seconds from the network deployment, and the last capture is simulated at 359,200 seconds from the network deployment ($12\cdot6$ hours later). These resulting in a total of 507 simulated captures for every combination of the protocol and $\lambda$ chosen. For every simulation, we run the protocol and measured the detection time and the number of messages sent from the nodes in the network.

In the simulation results, we count the number of messages sent by the nodes when a node-missing alarm or a presence-claim is broadcasted. Moreover, in the AdaBo protocol we count the messages sent when two nodes exchange their tokens, and the messages sent when a node asks for cooperation in the last small time-interval $\gamma$ of $\lambda$ before sending a capture alarm.
As we do not want that the set-up phase influences the evaluation of the general cost of the protocol (that in practice could last more that the simulated time), we start measuring the performance of the protocols from the 84,000th second onwards. That is, the messages sent before the 84,000th second are not considered in the final mean value of sent messages.

Note that we assume that a node becomes aware of the nodes that are in its communication range, thanks to the communication activity of the nodes, or thanks to the control messages of the network; as in the case of the INFOCOM traces, the bluetooth protocol. Thus, we do not consider  the communication activity for the meeting event as an overhead of the capture detection protocol.

As for the number of memory slots used by the protocols, we considered a small value because we observed that increasing the number of tracked nodes would just lead to an higher protocol overhead while not improving the protocol performance in terms of detection. This characterization is common for all the protocols we considered in this work. Furthermore, we observe that using a small value as for memory slots is particularly suitable for resource-constrained devices like sensor network.
In particular, we used one slot for the tracked node, and 5 slots for the SMSs of the Adaptive and AdaBo protocols.

In the simulation performed for the AdaBo protocol, the nodes start exchanging their token after 42,000 seconds from the network deployment. Furthermore, a node can propose a token exchange with another node if its exchange counter is not greater than 3. The time-interval $\gamma$ during which a node asks for cooperation before sending a node-missing alarm is equal to 3,600 seconds.

\subsection{Protocol Overall Performance}
\label{OVperformances}

\begin{figure}[hbt]
\begin{center}
	\includegraphics[scale=0.30]{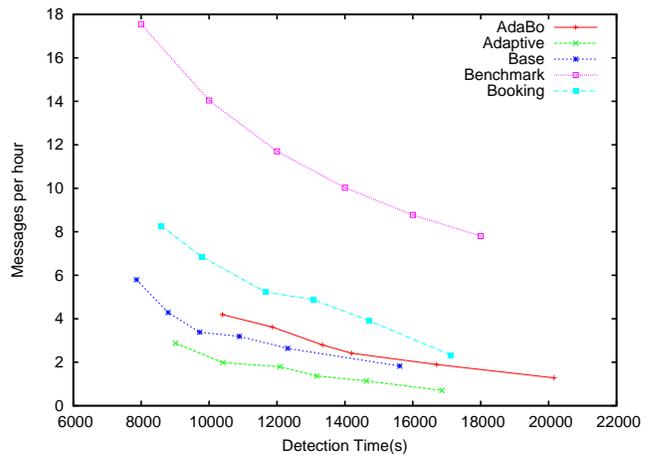}
\end{center}
\caption{INFOCOM traces: Cost vs. Detection}
\label{fig:infocom_detectTime_messages}
\end{figure}

In Figure~\ref{fig:infocom_detectTime_messages}, we plot the results of the simulations: the mean of the messages sent by each node per hour (y-axis) for multiple resulting detection times (x-axis). Each point is obtained as the mean of the results of 507 simulations executed for each specific protocol and a fixed $\lambda$. In particular, for each protocol we report the results for 6 different values of $\lambda$: 12,600, 14,400, 16,200, 18,000, 19,800, and 23,400 seconds.

We observe that the Benchmark protocol has the worst performance with respect to the other simulated protocols. In fact, chosen a desired detection time, it required the higher number of messages per hour. For example, given a detection time equals to 12,000 seconds, each node has to send almost 12 messages per hour. Meanwhile, the other protocols send about 5.2 messages per hour in the worst case (i.e. the Booking protocol). Note that the protocols we implemented, that are the AdaBo, the Adaptive, the Base, and the Booking protocol, all leverage the meeting events to detect a node capture. This confirms our intuition that mobility and
social ties can be leveraged to increase the performance of the protocols used in the network.

A fair comparison between the AdaBo, the Adaptive, the Base, and the Booking protocols can be made dividing these protocols in two classes:
\begin{itemize}
\item \textit{first class.} AdaBo and the Booking protocols.
\item \textit{second class.} Adaptive and the Base protocols.
\end{itemize}
The protocols in the first class guarantee that all the nodes of the network are monitored. The protocols in the second class do not guarantee that all the nodes of the network are monitored. When a protocol of the first class is used, we can always detect a node capture. While, when a protocol of the second class is used, it could be possible that a node capture goes undetected. That is, a false negative can occur. Because there is not a first class protocol that has better performance than the ones in the second class, there is not a protocol without false negative and with performance better than the others. Thus, assuming that the network administrator can not tolerate false negative detection, she has to adopt a protocol of the first class; otherwise, she can use a protocol of the second class, depending on the number of false negative the network can bear (discussed later). From the results reported in Figure~\ref{fig:infocom_detectTime_messages}, we can observe that the property of the first class protocols, that is to guarantee that all the nodes are tracked, comes at the prize of more messages sent. In fact, the AdaBo and the Booking protocols, on the contrary of the Adaptive and the Base protocol, monitor all the node in the network. Consequently, they monitor also the \textit{isolated} nodes, even if these isolated nodes cause a high number of node-missing alarm.

In the Booking protocol, we assume that the network administrator decides for every node what is the ID of the other node it is going to track. In particular, in our simulation the node with ID $i$ monitors the node with ID $i+1$ modulo 39. Figure~\ref{fig:infocom_detectTime_messages} also shows that between the two protocols of the first class, the AdaBo has better performance than the Booking one. We remind the reader that, the Booking protocol, differently from the AdaBo one, does not leverage the communities to optimize the node tracking. The results support our thesis that the social characteristics of the networks can be leveraged to increase the protocols performance.

Comparing the protocols of the second class, the Base and the Adaptive ones, the Adaptive is better than the Base protocol under two points of view. In the first place, from the point of view of the performance. In fact, in Figure~\ref{fig:infocom_detectTime_messages} we can see that, fixed a capture detection time, in the Adaptive protocol the number of messages sent per hour is lower than the ones in the Base protocol. In the second place, from the point of view of the false negative percentage. Indeed, from our simulation results, we observed that the mean of the false negative of the Base protocol over all the simulations is about 43.6\%, while the mean of the Adaptive protocol is about
42.9\%.
Thus, the Adaptive protocol not only has better performance respect to the Base one, but it has also a lower percentage of false negative.

\section{Conclusions}
\label{conclusions}

In this paper we have showed that it is possible to leverage both node mobility and communities of nodes that naturally emerge in mobile networks to enforce security properties. In particular, we have designed two class of protocols that take into consideration realistic mobility models to thwart node capture attack.
The first class of protocols provides the monitoring of the whole nodes in the network, sacrificing some efficiency,
while the second one releases the control on isolated nodes, achieving efficiency gains.

The proposed protocols have been tested on real traces, and the results confirmed our intuition: protocols leveraging emergent social ties provide better performances
than protocols leveraging mobility only.
To the best of our knowledge, this is the first result in the area and could open up a vein of research aimed at combining mobility and emergent social ties in mobile networks to enforce  security properties.


\balance

\bibliographystyle{abbrv}

\bibliography{mybib}

\newpage

\end{document}